\documentclass[a4paper,12pt]{article}
\usepackage{latexsym}
\usepackage[colorlinks]{hyperref}
 \makeatletter

\def\be{\begin{equation}}
\def\ee{\end{equation}}
\def\ba{\begin{array}} \def\ea{\end{array}}
\def\bea{\begin{eqnarray}}
\def\eea{\end{eqnarray}}

\def\R{\mathbb R}

\def\N{\mathbb N}
\def\Z{\mathbb Z}
\usepackage{amsmath,amssymb}

\makeatletter
 \@addtoreset{equation}{section}

 \makeatother

\begin{document}
\thispagestyle{empty}

\vspace{20pt}

\baselineskip15pt

\begin{center}
{\Large \Large \bf Nonlinear supersymmetry:\\
\vskip 0.2cm
from classical to
quantum mechanics
}

\vspace{10mm}

Mikhail Plyushchay\footnote{e-mail: mplyushc@lauca.usach.cl}
$^{a,\,b}$

\vspace{5mm}
 ${}^a$ Departamento de F\'{\i}sica,
Universidad de Santiago de Chile,
Casilla 307, Santiago 2, Chile

\vspace{5mm}
${}^b$ Institute for High Energy Physics,
Protvino, Russia

\end{center}

\vspace{5mm}
\begin{abstract}
Quantization of the nonlinear supersymmetry faces a problem
of a quantum anomaly. For some classes of superpotentials,
the integrals of motion admit the corrections guaranteeing
the preservation of the nonlinear supersymmetry
at the quantum level. With an example of the system
realizing the nonlinear superconformal symmetry, we discuss
the nature of such corrections and speculate on their
possible general origin.
\end{abstract}

\newpage

\section{Introduction}
Nonlinear symmetry is characterized by the
polynomial commutation relations of the integrals of motion
\cite{Walg}.
Nonlinear generalization of the usual supersymmetry
\cite{SW}
was obtained originally by employing the higher derivative
supercharges \cite{A1}.
The higher derivative realization
of the nonlinear supersymmetry was studied
in various aspects, see refs.
 \cite{A1}--\cite{F2}.

Another generalization of supersymmetric quantum mechanics
is a construction of the minimally bosonized supersymmetry
\cite{BoS}, in which a  reflection operator is employed
for
the role of the grading operator.
The development of the construction
in application to the study of hidden symmetries
of the purely parabosonic systems resulted in
observation of the nonlinear supersymmetry
independent from the original line
of research \cite{P1}.

A new wave of interest to the nonlinear supersymmetry
\cite{KP1}--\cite{Tan2} was triggered by the
observation of its close relation to the quasi-exactly
solvable systems
\cite{TurU}. The relationship was revealed, in particular,
under the study of the quantum anomaly problem \cite{KP1}
arising in quantization of
classical systems possessing nonlinear  supersymmetry
\cite{P1}.

In the present paper, we discuss
the solution of the quantum anomaly  problem associated
with the nonlinear supersymmetry
making emphasis on the analysis
of the nature of the corresponding quantum corrections.

The layout of the paper is as follows.
In Section 2, the origin of the quantum anomaly
problem and its  solution resulting in
the holomorphic form of the nonlinear supersymmetry
are discussed together with a universal algebraic
structure underlying the latter.
In Section 3 we consider a particular case of the
second order supersymmetry given by the Calogero-like
quantum Hamiltonian.
Section 4 is devoted to the discussion of the nonlinear
generalization of superconformal symmetry
and its interpretation from the point of view of
the special reduction procedure applied to the
associated free particle system.
In Section 5 we specify some problems
to be interesting for further consideration.

\section{Quantum anomaly and nonlinear holomorphic
supersymmetry}
Let us consider a system described by the
Lagrangian
\begin{equation}\label{Ln}
L=\frac{1}{2}(\dot{x}^2-W^2(x))
-gW'(x)\theta^+\theta^-+i\theta^+\dot{\theta}{}^-
\end{equation}
with Grassmann variables $\theta^\pm$,
$(\theta^+)^*=\theta^-$,
a superpotential $W(x)$ and a
`boson-fermion coupling constant' $g$.
At $|g|=1$
the pseudo-classical system (\ref{Ln}) underlies
Witten supersymmetric quantum mechanics with associated
Lie superalgebra of the integrals of motion,
while at $|g|=n$, $n=2,3,...$, it
possesses the nonlinear supersymmetry
\cite{P1,KP1}.
This can easily be seen in the canonical approach
where the Hamiltonian
\begin{equation}\label{Hn}
H=\frac{1}{2}(p^2+W^2(x))
+n W'(x)\theta^+\theta^-, \qquad n\in\Z,
\end{equation}
and the Poisson-Dirac brackets
$\{x,p\}=1$, $\{\theta^+,\theta^-\}=-i$
correspond to Lagrangian (\ref{Ln})
with $g=n$.
Due to the invariance of the Lagrangian under
$g\rightarrow -g$, $W(x)\rightarrow -W(x)$,
we shall imply $n\in \Z_+$ in what follows.
The system (\ref{Hn}) possesses the two odd integrals
of motion
\begin{equation}\label{Qn}
Q^+=z^n\theta^+,\qquad Q^-=(Q^+)^*=\bar{z}{}^n\theta^-
\end{equation}
in addition to the Hamiltonian which in terms  of the
complex
variables
\begin{equation}\label{z}
z=W(x)+ip,\qquad \bar{z}=z^*
\end{equation}
is represented in the form
\begin{equation}\label{Hholc}
H=\frac{1}{2}\left(z\bar{z}+in\{z,\bar{z}\}\theta^+\theta^-
\right).
\end{equation}
For integer values of the coupling constant, $g=n$, the
projections on the unit
of the Grassmann algebra of the
instant  frequencies of the oscillator-like
bosonic, $z$ ($\bar{z}$), and fermionic, $\theta^+$
($\theta^-$)
variables are commensurable (being equal
to $-W'(x)$ and $nW'(x)$
for $z$ and $\theta^+$, respectively),
that guarantees the existence
of the
local in time odd integrals of motion (\ref{Qn})
\cite{P1,KP1}.

The integrals satisfy the relations
\begin{equation}\label{susyn}
\{Q^+,Q^-\}=-iH^n,\quad
\{Q^+,Q^+\}=\{Q^-,Q^-\}=0,\quad \{Q^\pm,H\}=0.
\end{equation}
The case $n=0$ is characterized by
the trivial odd integrals of motion, $Q^\pm=\theta^\pm$.
The case $n=1$ corresponds to the linear supersymmetry, for
which
the supercharges and the Hamiltonian form
a classical Lie superalgebra corresponding to Witten
supersymmetric
quantum mechanics.
For $n=2,3,...$ the system possesses the nonlinear
supersymmetry
characterized by the nonlinear (polynomial of the order $n$)
Poisson superalgebra of the integrals of motion.

The essential difference of the nonlinear supersymmetry from
the
linear one appears as soon as we try to quantize the system.
Taking the direct quantum analogs for the supercharges,
\begin{equation}\label{qQn}
\hat{Q}{}^+=(\hat{Q}{}^-)^\dagger=\hat{z}{}^n\hat{\theta}{}^
+\qquad
{\rm with}\qquad  \hat{z}=W(x)+\hbar\frac{d}{dx},\qquad
\hat{\theta}{}^+=\sqrt{\hbar}\sigma_+,
\end{equation}
where $\sigma_+=\frac{1}{2}(\sigma_1+i\sigma_2)$,
and for the Hamiltonian,
\begin{equation}\label{qHn}
\hat{H}=\frac{1}{2}
\left(-\hbar^2\frac{d^2}{dx^2}+
W^2(x)+n\hbar W'(x)\sigma_3\right),
\end{equation}
one finds that at $n=1$ the quantum  supercharges
(\ref{qQn})
are the integrals of motion, $[\hat{Q}{}^\pm,\hat{H}]=0$,
for any choice of the superpotential $W(x)$, and the
superalgebra they
form
together with the Hamiltonian is a direct quantum analog
of (\ref{susyn}) with $n=1$.
However, for $n\geq 2$, the odd operators (\ref{qQn})
commute with the Hamiltonian (\ref{qHn}) only
for the quadratic superpotential
\begin{equation}\label{W2}
W(x)=w_2x^2+w_1x+w_0.
\end{equation}
In this case the classical nonlinear superalgebra
(\ref{susyn}) changes its form for
$[\hat{Q}{}^+,\hat{Q}{}^-]_{{}_+}=\hat{H}{}^n+P_{n-1}
(\hat{H})$,
where $P_{n-1}(\hat{H})$ is a polynomial of the order $n-1$
with the coefficients given in terms of $w_1^2-4w_0w_2$
and $w_2^2$, and disappearing for $\hbar\rightarrow 0$
\cite{KP1}.
For the superpotential of a generic form,
$[\hat{Q}{}^\pm,\hat{H}]\neq 0$
and we face the problem of the quantum anomaly.

In order to resolve this problem, one could look for
the class of the quantum analogs of the classical
supercharges
(\ref{Qn}) whose conservation would be consistent with
the quantum Hamiltonian of the fixed form (\ref{qHn}).
Then, one of the possibilities is to suppose that
the supercharge $\hat{Q}{}^+$ has a holomorphic form
depending only on
$\hat{z}$ (and not depending on $\hat{\bar{z}}$),
being a polynomial of the order $n$ in it.
The condition of conservation for such a supercharge
is reduced  to the
equation
\begin{equation}\label{difW}
n(n^2-1)\hbar\frac{d}{dx}(\hbar^2W''-\omega^2W)=0,
\end{equation}
where $\omega^2$ is a constant.
For $\omega\neq 0$,
the general solution to the equation (\ref{difW})
gives a three-parametric class of solutions
\begin{equation}\label{Wexp}
W(x)=w_+e^{\omega x}+w_-e^{-\omega x}+w_0,
\end{equation}
while for $\omega=0$ the solution is
(\ref{W2}).
Different choices for the parameters
$\omega$, $\omega_\pm$ and $\omega_0$
give rise then to a broad class of quasi-exactly solvable
and exactly solvable quantum mechanical systems \cite{KP1}.

For a quasi-exactly solvable Hamiltonian
only a finite number
of the  eigenstates and eigenvalues can be found
in a purely algebraic manner \cite{TurU}.
A finite-dimensional subspace spanned
by such eigenstates  provides with a
finite-dimensional representation of the $sl(2,R)$
algebra, and the dimension of the representation is
determined
by a natural parameter appearing
explicitly in the Hamiltonian.
In the case of the nonlinear supersymmetry of the
order $n$, the zero mode subspaces of supercharges
can be associated with the $n$-dimensional representations
of the $sl(2,R)$ \cite{KP1}.
In particular, the systems with nonlinear
supersymmetry corresponding to the quadratic superpotential
(\ref{W2}) can be related  \cite{KP3} to the quasi-exactly
solvable
systems with quartic potential appearing in the context of
the
PT-symmetric quantum mechanics \cite{Ben}.

The described class of the systems with nonlinear
supersymmetry
given by the holomorphic supercharges
corresponding to the superpotentials
(\ref{Wexp}) and (\ref{W2})
admits the following generalization.
First, we note that
$[\hat{z},\hat{\bar{z}}]=2\hbar W'(x)$ and
$[\hat{z},\hat{\bar{z}}]_{{}_+}=2(W^2(x)
-\hbar^2\frac{d^2}{dx^2})$.
Then, generalizing the operators $\hat{z}$ and
$\hat{\bar{z}}$
defined by Eq. (\ref{qQn}) for the pair of
mutually conjugate operators of an arbitrary nature,
$Z$ and  $\bar{Z}=Z^\dagger$,
we can change the Hamiltonian and equation (\ref{difW})
for
\begin{equation}\label{Hhol}
{\cal H}_n=\frac{1}{4}\left([Z,\bar{Z}]_{{}_+}+n[Z,\bar{Z}]
\sigma_3\right),
\end{equation}
\begin{equation}\label{DG}
[Z,[Z,[Z,\bar{Z}]]]=\omega^2[Z,\bar{Z}],\qquad
[\bar{Z},[\bar{Z},[Z,\bar{Z}]]]=\bar{\omega}{}^2[Z,\bar{Z}],
\end{equation}
where $\bar{\omega}=\omega^*$.
The relations (\ref{Hhol})
guarantee the existence of the two
integrals of motion for the system (\ref{Hhol}),
\begin{equation}\label{Qhol}
{\cal Q}^+_n=\prod_{k=0}^{n-1}\left(Z+\left(k-\frac{n-1}{2}
\right)\omega\right)\cdot
\sigma_+,\qquad
{\cal Q}^-_n=({\cal Q}^+)^\dagger,
\end{equation}
whose   structure
is corrected in comparison with the classical
case by a trivial shift of bosonic factors.

The nonlinear algebraic  relations similar to
relations (\ref{DG})
appeared also in the context of
spin integrable systems
under the name of the Dolan-Grady relations \cite{DG}.
Let us stress that here
the Dolan-Grady relations arise as necessary conditions
for anomaly-free quantization of pseudo-classical systems
with nonlinear holomorphic supersymmetry.

Through these relations, the Onsager algebra can be
generated \cite{KP3}. The case $\omega=0$
corresponds to a more simple form of the contracted
Onsager algebra generated recursively
by $Z_0\equiv Z$ and $\bar{Z}_0\equiv \bar{Z}$
via the contracted ($\omega=0$) Dolan-Grady relations:
\begin{eqnarray}
&[Z_m, \bar{Z}_n]=B_{m+n+1},\quad [Z_n,B_m]=Z_{m+n},\quad
[B_m,\bar{Z}_n]=\bar{Z}_{m+n},&\nonumber\\
\label{cOA}&[Z_m,Z_n]=0,\quad [\bar{Z}_m,\bar{Z}_n]=0,\quad
[B_m,B_n]=0,&
\end{eqnarray}
where $m,n\in\Z_+$ and $B_0=0$ is implied.
In general case of $\omega\neq 0$, the Onsager algebra
has a similar but more complicated form, see ref.
\cite{KP3}.
In both cases of $\omega=0$ and $\omega\neq 0$,
the Onsager algebra admits an infinite set of mutually
commuting
quadratic operators,
$$
J^l_\lambda=\frac
12\sum_{p=1}{l}\left([\bar{Z}_{p-1},Z_{l-p}]_{{}_+}
-B_lB_{l-p}\right)-\frac{\lambda}{2} B_l,
$$
where $l=1,2,\ldots$, and
$\lambda$ is a parameter. Putting $\lambda=n\in \N$ and
introducing the operators
${\cal J}_n^l=J_n^l\sigma_-\sigma_+
+J^l_{-n}\sigma_+\sigma_-$,
$\sigma_-=\sigma_+^\dagger,$
one finds that the Hamiltonian (\ref{Hhol}) belongs to the
infinite set of mutually commuting even operators
${\cal J}_n^l$, $l=1,2,\ldots$, ${\cal J}_n^1={\cal H}_n$,
to be quadratic in the generators of the Onsager algebra,
and that the nonlinear superalgebra takes the form
\begin{equation}\label{Alhol}
[{\cal Q}^+_n,{\cal Q}^-_n]_{{}_+}={\cal H}_n^n+P_{n-1}({
\cal J}_n^l),
\qquad
[{\cal Q}^\pm,{\cal J}^l_n]=[{\cal J}_n^l,{\cal J}_n^k]=0,
\end{equation}
where $P_{n-1}({\cal J}_n^l)$ is a polynomial
of the order $n-1$ \cite{KP3}.
For the systems with finite number of degrees of
freedom, there is only a finite number of
independent integrals
${\cal J}_n^l$.

\section{Second order supersymmetric quantum mechanics}
Though a universal algebraic structure
associated with Dolan-Grady relations
allows ones to realize nonlinear holomorphic supersymmetry
in nontrivial (Riemann)
and noncommutative  geometries \cite{KP4}
as well as to generalize the construction
for the case of nonlinear pseudo-supersymmetry \cite{KP3}
related to the $PT$-symmetric quantum mechanics
\cite{Most},
it generates a rather special class
of the systems possessing nonlinear supersymmetry.
But since there is no one-to-one correspondence between
the classical canonical and the quantum unitary
transformations,
one can look for other forms of nonlinear supersymmetry
proceeding from the classical representations
different from the holomorphic one.
For instance, one could look for the classical
formulation characterized by the supercharges to be
polynomials of the $n$-th degree in the momentum $p$.
The problem of finding such a formulation can be
solved completely in the simplest case
$n=2$ \cite{KP1}. For it, the
Hamiltonian and the supercharges
are fixed as
\begin{equation}\label{H2}
H=\frac{1}{2}\left(
p^2+W^2(x)-\frac{\nu}{W^2(x)}\right)
+2W'(x)\theta^+\theta^-, \qquad \nu\in\R,
\end{equation}
\begin{equation}\label{Q2}
Q^\pm=\frac{1}{2}\left((\pm ip+W(x))^2+\frac{\nu}{W^2(x)}
\right)\theta^\pm.
\end{equation}
The system (\ref{H2}), (\ref{Q2})
is characterized by the
order $n=2$ superalgebra
$\{Q^+,Q^-\}=-i(H^2+\nu)$.
Note that for $W(x)=x$, the part of
(\ref{H2}) without the last nilpotent term
takes the form of the Hamiltonian for the two-body Calogero
problem.
The peculiarity of the
Calogero-like $n=2$
supersymmetric system (\ref{H2}), (\ref{Q2})
in comparison with the
nonlinear holomorphic supersymmetry
is that it admits the quantum anomaly-free
formulation for the superpotential $W(x)$
of an arbitrary
form. The anomaly-free quantum version
of the $n=2$ supersymmetric system (\ref{H2}), (\ref{Q2}) is
given by the operators
\begin{equation}\label{H2q}
\hat{H}=\frac{1}{2}\left(-\hbar^2\frac{d^2}{dx^2}+W^2-
\frac{\nu}{W^2}
+2\hbar W'\sigma_3+\Delta(W)\right),
\end{equation}
\begin{equation}\label{Q2q}
\hat{Q}{}^+=(\hat{Q}{}^-)^\dagger=
\frac{1}{2}\left(\left(\hbar\frac{d}{dx}+W\right)^2+
\frac{\nu}{W^2}-\Delta(W)\right)\hat{\theta}{}^+,
\end{equation}
with the quantum correction
\begin{equation}\label{d2q}
\Delta(W)=\frac{\hbar^2}{4W^2}(2W''
W-W'{}^2)=\hbar^2\frac{1}{\sqrt{W}}(\sqrt{W})''.
\end{equation}
Note that here the quadratic and exponential forms of the
superpotential play also a special role:
for $W(x)=(w_1x+w_0)^2$, $\Delta(W)=0$,
while for $W(x)=(w_+e^{\omega x}+w_-e^{-\omega x})^2$,
$\Delta(W)=\hbar^2 \omega^2=const$.

The inclusion of the quantum correction in the Hamiltonian
and supercharges is
crucial for preservation of the nonlinear supersymmetry:
without it the supercharges would not be the
quantum integrals of motion.
The quantum integrals (\ref{H2q}) and
(\ref{Q2q})
satisfy the relation to be the direct quantum analog
of the corresponding classical relation:
$[\hat{Q}{}^+,\hat{Q}{}^-]_{{}_+}=\hat{H}{}^2+\nu$.

The quantum systems (\ref{H2q}), (\ref{Q2q})
were the first ones
discussed in the context
of the nonlinear ($n=2$) supersymmetry
\cite{A1}.
But it is interesting to
note that the Hamiltonian structure similar
to (\ref{H2q})
was discussed earlier in the context of partial symmetry
breaking in $N=4$ supersymmetric quantum mechanics
\cite{IKP}, and that the quantum term
(\ref{d2q}) appeared also in the method
of constructing new solvable potentials via the
operator transformations, see \cite{SW} and references
therein.
The order $n=2$ nonlinear supersymmetry given by
Eqs. (\ref{H2q}), (\ref{Q2q})
with no restrictions
on the form of the superpotential admits
a generalization for $n>2$ for a
class of superpotentials of a special form
related to some quasi-exactly solvable systems
\cite{Ao1}.

\section{Nonlinear superconformal symmetry and reduction}
The natural question arising here is whether the
origin of the quantum corrections appearing under
construction of nonlinear supersymmetry can be
explained to some extent? To clarify this point,
we consider a class of the systems
\cite{LP1,AnPl} possessing nonlinear
supersymmetry
and related to superconformal
mechanics model \cite{AP}. As we shall see, in the simplest
case $n=2$, the corresponding quantum system is
a particular case of the $n=2$ supersymmetric systems
(\ref{H2q}), (\ref{Q2q}),
and that the corresponding quantum corrections
necessary for conservation of the nonlinear supersymmetry
appear for it in a natural way via a
special reduction procedure applied to
the associated free particle spinning system.

So, let us consider a classical system
given by the Hamiltonian (\ref{Hn}) with the
superpotential $W(x)=\alpha x^{-1}$:
\begin{equation}\label{scH}
H=\frac{1}{2}\left(p^2+\alpha(\alpha-n\theta^+\theta^-)x^{-2
}\right).
\end{equation}
At $n=1$, Eq. (\ref{scH})
is  the Hamiltonian of  the superconformal mechanics
model \cite{AP}, which possesses more broad
superconformal symmetry generated by the
Hamiltonian and the three even integrals,
\begin{equation}\label{DKS}
D=\frac{1}{2}xp-tH,\quad
K=\frac{1}{2}x^2 -2tD-t^2H,\quad
\Sigma=\theta^+\theta^-,
\end{equation}
as well as by the four odd integrals of motion
$$
Q_a=p\theta_a+\alpha x^{-1}\epsilon_{ab}\theta_b,\quad
S_a=p\theta_a-tQ_a,
$$
where $a,b=1,2$,
and $\theta^\pm=\frac{1}{\sqrt{2}}(\theta_1\pm i\theta_2)$.
All these integrals satisfy the conservation
equation of the form
$\frac{d}{dt}I=\frac{\partial}{\partial t}
I+\{I,H\}=0$.
In the case of $n=2,3,\ldots$, the set of even integrals of
motion
is the same as in the superconformal mechanics model ($n=1$)
with the correspondingly changed boson-fermion coupling
constant
in the Hamiltonian, but instead of the 4 odd integrals of
motion,
there are $2(n+1)$ supercharges
\begin{equation}\label{Snl}
S^+_{n,l}=(x+itz)^lz^{n-l}\theta^+,\quad
S^-_{n,l}=(S^+_{n,l})^*,\quad
l=0,1,\ldots,n,
\end{equation}
where the even complex variable $z$ is
defined by Eq. (\ref{z}) with $W(x)=\alpha x^{-1}$.
These integrals form the following nonlinear
superconformal algebra $osp(2|2)_n$ \cite{LP1,AnPl}:
\begin{eqnarray}\label{hdk}
&\{D,H\}=H,\qquad
\{D,K\}=-K,\qquad
\{K,H\}=2D,&\\
&\{D,S^\pm\}=\left(\frac{n}{2}-l\right)S^\pm_{n,l},\qquad
\{\Sigma,S^\pm_{n,l}\}=\mp iS^\pm_{n,l},&\label{sc1}\\
&\{H,S^\pm_{n,l}\}=\pm ilS^\pm_{n,l-1},\qquad
\{K,S^\pm_{n,l}\}=\pm i(n-l)S^\pm_{n,l+1},&\label{sc2}\\
&\{S^+_{n,m},S^-_{n,l}\}=-i(2H)^{n-m-1}
(2K)^{l-1}\alpha_D^{m-l}\left[4HK -i\Sigma
(n(m-l)\alpha_D+4\alpha l(n-m))\right],&
\label{sc3}
\end{eqnarray}
where in the last relation $\alpha_D\equiv \alpha-2iD$
and  $m\geq l$, and the brackets between the odd integrals
for $m<l$ are obtained from it by the complex conjugation.
The nonlinear superconformal algebra
$osp(2|2)_n$ contains the bosonic
Lie subalgebra $so(1,2)\oplus u(1)$ generated by the even
integrals, while the set of the odd supercharges constitute
the two spin-$\frac{n}{2}$ representations of the bosonic
subalgebra.
At $n=1$ the nonlinear superconformal algebra
is reduced to the Lie superalgebra $osp(2|2)$.

The nonlinear superconformal symmetry admits the anomaly
free
quantization.
The corresponding quantum corrections to the integrals of
motion can be found directly giving the following quantum
analogs
of the $osp(2|2)_n$ generators:
\begin{equation}\label{Hscq}
\hat{H}=\frac{1}{2}\left(-\hbar^2\frac{d^2}{dx^2}+(a_n+b_n
\hbar \sigma_3)\frac{1}{x^2}\right),
\end{equation}
\begin{equation}\label{DKscq}
\hat{D}=-\frac{i}{2}\hbar\left(x\frac{d}{dx}+\frac 12\right)
-\hat{H}t,\quad
\hat{K}=\frac 12 x^2-2\hat{D}t-\hat{H}t^2,\quad
\hat{\Sigma}=\frac{\hbar}{2}\sigma_3,
\end{equation}
\begin{equation}\label{Sscq}
\hat{S}{}^+_{n,l}=(x+it{\cal D}_{\alpha-n+1})
(x+it{\cal D}_{\alpha-n+2})\ldots
(x+it{\cal D}_{\alpha-n+l}){\cal D}_{\alpha-n+l+1}
\ldots
{\cal D}_\alpha\sigma_+,\quad
\hat{S}{}^-_{n,l}=(\hat{S}{}^+_{n,l})^\dagger,
\end{equation}
where
\begin{equation}\label{abq}
a_n=\alpha_n^2+\frac{\hbar^2}{4}(n^2-1),\quad
b_n=-n\alpha_n,\quad
\alpha_n=\alpha-\frac{\hbar}{2}(n-1),\quad
{\cal D}_{\alpha-k}=\left(\hbar\frac{d}{dx}+
\frac{\alpha}{x}\right)-\hbar\frac{k}{x}.
\end{equation}
The second terms in $a_n$, $\alpha_n$ and
${\cal D}_{\alpha-k}$
are the quantum corrections which guarantee the conservation
of the quantum analogs of the integrals of motion
(\ref{scH}), (\ref{DKS}) and (\ref{Snl}).
We note that since the
not depending explicitly on time
supercharges $S^+_{n,0}=z^n\theta^+$ and
$S^-_{n,0}=(S^+_{n,0})^*$
have the holomorphic form
and the Hamiltonian (\ref{scH})
can be represented in the form (\ref{Hholc}),
classically the system (\ref{scH})
is the system with the nonlinear holomorphic supersymmetry
of the order $n$.
Furthermore, the quantum shift of the parameters
in the bosonic factors of the odd integrals (\ref{Sscq})
is similar to the shift in the factors of the
holomorphic supercharges (\ref{Qhol}).
However, the
corresponding superpotential $W=\alpha x^{-1}$
does not satisfy the equation
(\ref{difW}), and the form of the
quantum analogs of the Hamiltonian and supercharges
$S^\pm_{n,0}$ does not correspond
to that of the nonlinear holomorphic supersymmetry.
Note also that at $n=2$ the quantum Hamiltonian
$\hat{H}$  and the supercharges
$\hat{S}{}^\pm_{n,0}$ given by Eqs. (\ref{Hscq}),
(\ref{Sscq}) and (\ref{abq})
coincide with the Hamiltonian and supercharges
(\ref{H2q}), (\ref{Q2q})
of the system
with $W(x)=(\alpha-\frac{\hbar}{2})x^{-1}$ and $\nu=0$.

The structure of the quantum system
(\ref{Hscq})--(\ref{abq}) admits an interesting
interpretation.
Let us consider the system of a free planar spinning
particle
described by the Lagrangian
\begin{equation}\label{Lf}
L=\frac{1}{2}\dot{x}_i^2-\frac{i}{2}\dot{\xi}_i\xi_i.
\end{equation}
Being a free system, it is characterized by the integrals
of motion $p_i$, $X_i=x_i-p_it$ and $\xi_i$.
Via the Poisson-Dirac brackets $\{x_i,p_j\}=\delta_{ij},$
$\{\xi_i,\xi_j\}=-i\delta_{ij}$, $i,j=1,2$,
they generate the space translations, the Galilei boosts and
the supertranslations.
Any function of $p_i$, $X_i$ and $\xi_i$ is also an integral
of motion,
and, in particular, the even quadratic functions
\begin{equation}
H=\frac{1}{2}p_i^2,\quad
K=\frac{1}{2}X_i^2,\quad
D=\frac{1}{2}X_ip_i,\quad
\Sigma=-\frac{i}{2}\epsilon_{jk}\xi_j\xi_k,
\end{equation}
\begin{equation}
J=L+\Sigma, 
\end{equation}
where $L=\epsilon_{ij}X_ip_j,$
are the integrals of motion
generating the Lie algebra $so(1,2)\oplus u(1)\oplus u(1)$.
Here the two terms $u(1)$ correspond to the spin $\Sigma$
and the total angular momentum $J$, while the conformal
symmetry $so(1,2)$ is generated by the
Hamiltonian $H$, by the scale transformations generator $D$,
and by the
generator of the special conformal transformations $K$.

The system (\ref{Lf}) has more bosonic degrees of freedom
than the system (\ref{scH}),
and bosonic and fermionic degrees
of freedom are not coupled in it.
One can decrease the number of variables
and  couple appropriately the bosonic and fermionic degrees
of freedom
by introducing the constraint fixing the orbital motion:
\begin{equation}\label{Jn}
{\cal J}_n-\alpha=0,\qquad
{\cal J}_n\equiv L+n\Sigma,
\end{equation}
where $\alpha$ is a real constant and $n\in \N$.
Then the quantities having zero brackets with the constraint
(\ref{Jn})
are invariant with respect to the gauge transformations
generated by it, and can be identified as
observables of the system (\ref{Lf})
supplemented with the constraint (\ref{Jn}).
In order to identify all the set of observables,
we define the complex variables
\begin{equation}
X_\pm=\frac{1}{\sqrt{2}}(X_1\pm i X_2),\quad
P_\pm=\frac{1}{\sqrt{2}}(p_1\pm i p_2),\quad
\xi_\pm=\frac{1}{\sqrt{2}}(\xi_1\pm i\xi_2),
\end{equation}
with nontrivial brackets $\{X_+,P_-\}=\{X_-,P_+\}=1$,
$\{\xi_+,\xi_-\}=-i$,
and find that in addition to the even observables
$$
H=P_+P_-,\quad
D=\frac{1}{2}(X_+P_-+P_+X_-),\quad
K=X_+X_-,\quad
\Sigma=\xi_+\xi_-
$$
and ${\cal J}_n$ given by Eq. (\ref{Jn}) with
$L=i(X_+P_--X_+P_-)$,
there is a set of odd observables
\begin{equation}
S^+_{n,l}=2^{n/2}i^{n-l}P_-^{n-l}X_-^l\xi_+,\qquad
S^-_{n,l}=(S^+_{n,l})^*.
\end{equation}
This set of even and odd observables forms on the constraint
surface
the Lie superalgebra $osp(2|2)$ ($n=1$),
or the nonlinear superconformal algebra $osp(2|2)_n$ ($n>1)$
given by Eqs.
(\ref{hdk})--(\ref{sc3}).
The free planar spinning particle system  being
reduced classically to the constraint surface reproduces
the system (\ref{scH}).
Moreover, the application of the Dirac quantization
prescription
applied to the system (\ref{Lf}), (\ref{Jn})
(first quantize and then reduce),
reproduces correctly all the quantum corrections
to the integrals of motion
necessary for preservation of the nonlinear superconformal
symmetry
at the quantum level (for the details see ref. \cite{AnPl}).

\section{Discussion and outlook}
We have seen that
the quantization of the nonlinear supersymmetry faces
the problem of the quantum anomaly. For some classes of the
superpotentials, it is possible to find the corrections
to the integrals of motion
which guarantee the preservation of the nonlinear
supersymmetry
at the quantum level.  However the origin
of such corrections  is not clear in a generic case.

The example of the system (\ref{scH})
shows that the corrections
can be understood as appearing due to
the quantum Dirac reduction procedure applied
to the appropriately chosen free extended system
supplied with the constraint
generating the necessary boson-fermion coupling.
It was also observed in \cite{KP2} that the nonlinear
supersymmetry
of the non holomorphic form \cite{Ao1} related to
quasi-exactly solvable systems
with sextic potential can be obtained via reduction of
a planar system possessing nonlinear
holomorphic supersymmetry.
At the same time, it is known that a broad class of the
quasi-exactly
solvable systems, to which the systems with nonlinear
supersymmetry
are intimately related,
can be obtained by the dimensional
reduction of the corresponding extended systems \cite{Tur2}.
Note that the same is true with respect to some integrable
systems
\cite{OlP}.
The nonlinear symmetries of a non-supersymmetric nature
can also be obtained by the Hamiltonian reduction
\cite{Walg}.

Therefore, it would be interesting to investigate the
problem
of the quantum anomaly associated with nonlinear
supersymmetry
within the general framework of reduction.
This could be helpful, in particular, for establishing
the nature of the universal quantum
correction (\ref{d2q}) for the Calogero-like systems
with the $n=2$ supersymmetry.
The idea of reduction could  also be useful
for clarifying the origin of the Dolan-Grady relations,
on which the construction of the nonlinear holomorphic
supersymmetry
is based, as well as for its generalization
for the case of many particle systems.
In such a way one could try to answer the
intriguing question on possibility to realize
the nonlinear supersymmetry at the field level.

\vskip 0.3cm

{\bf Acknowledgements.}
The author thanks the organizers of the
Conference ``Progress in supersymmetric quantum mechanics"
for the warm hospitality and stimulating
atmosphere.
This work was partially supported by the grant
1010073 from FONDECYT (Chile) and by DICYT-USACH.


\begin{thebibliography}{99}
\bibitem{Walg}
J. de Boer, F. Harmsze, T. Tjin,
Phys. Rept. {\bf 272} (1996) 139
[hep-th/9503161].

\bibitem{SW}
F. Cooper, A. Khare and U. Sukhatme,
Phys. Rep. {\bf 251} (1995) 267
[hep-th/9405029].


\bibitem{A1}
A.~A.~Andrianov, M.~V.~Ioffe and V.~P.~Spiridonov,
Phys.\ Lett.\ A {\bf 174} (1993) 273
[hep-th/9303005];\\
A.~A.~Andrianov, F.~Cannata, J.~P.~Dedonder and M.~V.~Ioffe,
Int.\ J.\ Mod.\ Phys.\ A {\bf 10} (1995) 2683
[hep-th/9404061];\\
A.~A.~Andrianov, M.~V.~Ioffe and D.~N.~Nishnianidze,
Phys.\ Lett.\ A {\bf 201} (1995) 103,
[hep-th/9404120].

\bibitem{S1}
V. G. Bagrov and B. F. Samsonov, Teor. Math. Phys.
{\bf 104} (1995) 1051;\\
V. G. Bagrov and B. F. Samsonov,
Phys.\ Part.\ Nucl.\  {\bf 28} (1997) 374;\\
B.~F.~Samsonov,
Mod.\ Phys.\ Lett.\ A {\bf 11} (1996) 1563
[quant-ph/9611012];\\
B.~F.~Samsonov,
Phys.\ Lett.\ A {\bf 263} (1999) 274
[quant-ph/9904009].

\bibitem{A4}
A.~A.~Andrianov, F.~Cannata, M.~V.~Ioffe and
D.~N.~Nishnianidze,
J.\ Phys.\ A {\bf 30} (1997) 5037
[quant-ph/9707004].

\bibitem{FH}
D. J. Fernandez  and V. Hussin, J. Phys. A {\bf 32} (1999)
3603.

\bibitem{Nie}
B. Mielnik, L.M. Nieto, and O. Rosas-Ortiz.
Phys. Lett. A {\bf 269} (2000) 70.



\bibitem{PlFM}
M.~S.~Plyushchay,
Phys.\ Lett.\ B {\bf 485} (2000) 187
[hep-th/0005122].


\bibitem{F2}
D.~J.~Fernandez, R.~Munoz and A.~Ramos,
Phys.\ Lett.\ A {\bf 308} (2003) 11
[quant-ph/0212026];\\
D.~J.~Fernandez C. and E.~Salinas-Hernandez,
J.\ Phys.\ A {\bf 36} (2003) 2537
[quant-ph/0303123].


\bibitem{BoS}
M.~S.~Plyushchay,
Annals Phys. (NY)  {\bf 245} (1996) 339
[hep-th/9601116];\\
M.~S.~Plyushchay,
Mod.\ Phys.\ Lett.\ A {\bf 11} (1996) 397
[hep-th/9601141];\\
J.~Gamboa, M.~Plyushchay and J.~Zanelli,
Nucl.\ Phys.\ B {\bf 543} (1999) 447
[hep-th/9808062].


\bibitem{P1}
M.~Plyushchay,
Int.\ J.\ Mod.\ Phys.\ A {\bf 15} (2000) 3679
[hep-th/9903130].

\bibitem{KP1}
S.~M.~Klishevich and M.~S.~Plyushchay,
Nucl.\ Phys.\ B {\bf 606} (2001) 583
[hep-th/0012023].

\bibitem{KP2}
S.~M.~Klishevich and M.~S.~Plyushchay,
Nucl.\ Phys.\ B {\bf 616} (2001) 403
[hep-th/0105135].


\bibitem{Ao1}
H. Aoyama, M. Sato and T. Tanaka,
Phys. Lett. B {\bf 503} (2001) 423-429,
[quant-ph/0012065];\\
H. Aoyama, M. Sato and T. Tanaka,
Nucl.\ Phys.\ B {\bf 619} (2001) 105-127
[quant-ph/0106037];\\
H. Aoyama, N. Nakayama, M. Sato and T. Tanaka,
Phys. Lett. B {\bf 519} (2001) 260
[hep-th/0107048].



\bibitem{dorey}
P. Dorey, C. Dunning and R. Tateo,
J. Phys. A {\bf 34} (2001) 5679
[hep-th/0103051].


\bibitem{KP3}
S.~M.~Klishevich and M.~S.~Plyushchay,
Nucl.\ Phys.\ B {\bf 628} (2002) 217
[hep-th/0112158].


\bibitem{KP4}
S.~M.~Klishevich and M.~S.~Plyushchay,
Nucl.\ Phys.\ B {\bf 640} (2002) 481
[hep-th/0202077];\\
S.~M.~Klishevich and M.~S.~Plyushchay,
J.\ Phys.\ A {\bf 36} (2003) 11299
[hep-th/0212117].

\bibitem{Tan1}
T.~Tanaka,
Phys.\ Lett.\ B {\bf 567} (2003) 100
[hep-th/0202101];\\
T.~Tanaka,
Annals Phys. (NY) {\bf 309} (2004) 239
[hep-th/0306174].

\bibitem{A5}
A.~A.~Andrianov and A.~V.~Sokolov,
Nucl.\ Phys.\ B {\bf 660} (2003) 25
[hep-th/0301062].



\bibitem{Tan2}
A.~Gonzalez-Lopez and T.~Tanaka,
``A new family of N-fold supersymmetry: Type B,''
[hep-th/0307094].


\bibitem{TurU}
A. Turbiner,
Comm. Math. Phys. {\bf 118} (1988) 467;\\
A. Ushveridze,
Quasi-exactly solvable models in quantum mechanics,
(IOP Publishing, Bristol, 1994).


\bibitem{Ben}
C. M. Bender and S. Boettcher,
J. Phys. A {\bf 27} (1998) L273
[physics/9801007].


\bibitem{DG}
L. Dolan and M. Grady,
Phys. Rev. D {\bf 25} (1982) 1587.



\bibitem{Most}
A.~Mostafazadeh,
J. Math. Phys. {\bf 43}  (2002) 205
[math-ph/0107001];\\
M.~Znojil, F.~Cannata, B.~Bagchi and R.~Roychoudhury,
Phys. Lett. B {\bf 483} (2000) 284
[hep-th/0003277].


\bibitem{IKP}
E.~A.~Ivanov, S.~O.~Krivonos and A.~I.~Pashnev,
Class.\ Quant.\ Grav.\  {\bf 8} (1991) 19.



\bibitem{LP1}
C.~Leiva and M.~S.~Plyushchay,
JHEP {\bf 0310} (2003) 069
[hep-th/0304257];\\
C.~Leiva and M.~S.~Plyushchay,
Phys. Lett. B {\bf 582} (2004) 135
[hep-th/0311150].


\bibitem{AnPl}
A.~Anabalon and M.~S.~Plyushchay,
Phys.\ Lett.\ B {\bf 572} (2003) 202
[hep-th/0306210].



\bibitem{AP}
V.~P.~Akulov and A.~I.~Pashnev,
Teor.\ Mat.\ Fiz.\  {\bf 56} (1983) 344;\\
S.~Fubini and E.~Rabinovici,
Nucl.\ Phys.\ B {\bf 245} (1984) 17.


\bibitem{Tur2}
A.~Turbiner,
``Quasiexactly solvable differential equations'',
in: CRC Handbook of Lie Group Analysis of
Differential Equations,
 Vol. 3: New Trends in Theoretical Developments and
 Computational Methods,
 ed. N.H. Ibragimov (CRC Press, 1995)
[hep-th/9409068].


\bibitem{OlP}
M.~A.~Olshanetsky and A.~M.~Perelomov,
Phys.\ Rept.\  {\bf 71} (1981) 313.



\end{thebibliography}
\end{document}